\documentclass[twocolumn,showpacs,prd,floatfix,axodraw]{revtex4}
\usepackage{mathrsfs}
\usepackage{graphicx,,booktabs,bm}
\usepackage{overpic}
\usepackage{color}
\usepackage{amssymb}

\usepackage{mathrsfs,bm,amsmath,amssymb}
\usepackage{longtable,lscape}
\usepackage{txfonts}
\usepackage{amssymb}
\usepackage{indentfirst}
\usepackage{graphicx,,booktabs}
\usepackage{multirow}
\usepackage{color}
\usepackage{amssymb}

\definecolor{cover}{rgb}{0.77,0.87,0.88}
\definecolor{blueone}{rgb}{0.1,0.1,.7}
\definecolor{citec}{rgb}{0.14,0.47,0.09}
\definecolor{two}{rgb}{0.0,0.5,0.}
\definecolor{three}{rgb}{.5,.1,0.15}
\usepackage[bookmarks=true,bookmarksopen=false,plainpages=false,breaklinks=true,
   bookmarksnumbered=true,hypertexnames=false,
   filecolor=blue,urlcolor=three,menucolor=three,
   linkcolor=three,citecolor=blueone, colorlinks,
   anchorcolor=blue,runcolor=pink,frenchlinks=red
   pdfstartview=FitH,pdftitle=title,%
   pdfauthor=author]{hyperref}

\begin{document}
\title{{{Photoproduction of possible pentaquark states $\Lambda^0_b(5912)$
and $\Lambda^0_b(5920)$ in the $\gamma{}p\to{}\Lambda_b^{0(*)}B^+$ reactions}}}

\author{Yin Huang$^1$}
\author{Hong Qiang Zhu$^{2}$}
\email{20132013@cqnu.edu.cn}

\affiliation{$^1$School of Physical Science and Technology, Southwest Jiaotong University, Chengdu 610031,China\\
$^2$College of Physics and Electronic Engineering, Chongqing Normal University, Chongqing 401331,China}

\date{\today}
\begin{abstract}
In this work,  we report on a theoretical study of possible pentaquark states $\Lambda^0_b(5912)$
and $\Lambda^0_b(5920)$ in the $\gamma{}p\to{}\Lambda_b^{0(*)}B^+$ reactions within an effective Lagrangian
approach.   In addition to the contributions from the $s$-channel nucleon pole and $t$-channel $\bar{B}^{{*}-}$
exchange, the contact term contribution are also included. Our theoretical approach is based on the chiral unitary
theory where the $\Lambda^0_b(5912)$ and $\Lambda^0_b(5920)$ resonances are dynamically generated.  Within the coupling
constants of the $\Lambda^0_b(5912)$ and $\Lambda^0_b(5920)$ to $\bar{B}p$ and $\bar{B}^{*}p$ channels
obtained from chiral unitary theory, the total and differential cross sections of the $\gamma{}p\to{}\Lambda_b^{0(*)}B^+$
are evaluated.  Our calculation indicates that the cross section for $\gamma{}p\to{}\Lambda_b^{0}(5912)B^{+}$ and
$\gamma{}p\to{}\Lambda_b^{0}(5920)B^{+}$ reactions are of the order of 0.0164 nb and 0.00527 nb,respectively.
If measured in future experiments, such as Electron-Ion Collider in China (EicC) or US(US-EIC), the predicted
total cross sections and specific features of the angular distributions can be used to test the (molecular) nature of the
$\Lambda^0_b(5912)$ and $\Lambda^0_b(5920)$ that they may be pentaquark states.
\end{abstract}


\maketitle
\section{INTRODUCTION}
Thanks to the experimental progress in the sector of heavy baryons in the past decade. Many heavy baryons that can not be ascribed into
3-quark(qqq) configurations have been reported~\cite{Zyla:2020zbs}.  Such as, three narrow hidden-charm pentaquarks, namely
$P_c(4312)$,$P_c(4440)$, and $P_c(4450)$, were observed by the LHCb Collaboration in the $J/\psi{}p$ invariant mass distributions of
the $\Lambda_b\to{J/\psi}pK$ decay~\cite{Aaij:2019vzc}.  Soon afterwards, a new possible strange hidden charm pentaquark $P_{cs}(4459)$
was observed by the LHCb Collaboration with $udsc\bar{c}$ conponenet in the $\Xi_b^{-}\to{}J/\psi{}\Lambda{}K^{-}$ process~\cite{Aaij:2020gdg}.
These findings intrigue an active discussion on the structure of these states.  A classical way to describe these states is treating
them as  candidates of meson-baryon molecular pictur~\cite{Chen:2020uif,Chen:2021tip}.

Investigating the pentaquark states has been a long history.  Even before the quark model was proposed by Gell-Mann and Zweig,
the $\Lambda(1405)$ had been suggested as a bound state from the $\bar{K}N$ interaction~\cite{Dalitz:1959dn}.  To understand
the strange magnetic momentum problem and the mass inversion problem,  Zou and his collaborators propose that there should exist
considerable five-quark($uuds\bar{s}$) configurations in the $N(1535)$~\cite{Zou:2010tc}, and the five-quark components also
provided a natural explanation for its large couplings to the strange $\bar{K}\Lambda$,$\bar{K}\Sigma$,$N\phi$,and $N\eta^{'}$
channels~\cite{Liu:2005pm}.  By analyzing the experimental data,  people suggested that $N(1875)$ and $N(2100)$ might be
hidden-strangeness pentaquark instead of naive three-quark state~\cite{He:2017aps}.  Even though there are many "missing resonances"
in this mass region that were predicted by the quark model.  The detailed discussions about the pentaquark states can be found
in Ref.~\cite{Guo:2017jvc}.

Although people interpret some states as the pentaquark states, other possible explanations such as the three quark state
(as long as quantum numbers allow, it might well be the case) can not be fully excluded.  The possible reason is that the quark pair
creation model still suffers from relatively large uncertainties~\cite{Liu:2009fe}.  Strictly speaking, these hadrons are not
perfect candidate of pentaquark states.
 We take the state $\Xi_b(6227)$ as an example.
Being considered as a traditional bottom baryon with $dsb$ three quark component, the strong decays of this state have been studied by using
the heavy quark light diquark model~\cite{Chen:2018orb} and QCD sum rule~\cite{Azizi:2020azq}.  On the other hand, many works
treat $\Xi_b(6227)$ as candidates of the molecular state with $\bar{u}udsb$ five quark component~\cite{Huang:2018bed, Yu:2018yxl}
because the mass gap between the $\Xi_b(6227)$ and the ground $\Xi_b$, about 440 MeV, is large enough to excite a light quark-antiquark
pair $\bar{u}u$ to form a molecular state.
 However, it is easy to confirm the new structures $P_c$($P_c(4312)$,$P_c(4440)$, and $P_c(4450)$)
and $P_{cs}(4459)$ contain at least five valence quarks according to the quark components of decay
model $J/\psi{p}$ and $J/\psi{}\Lambda$~\cite{Aaij:2019vzc,Aaij:2020gdg}. So they are perfect candidates of hidden-charm
pentaquark states.

However, in addition to finding the four hidden-charm pentaquark states in LHCb experiment~\cite{Aaij:2019vzc, Aaij:2020gdg}, there is no
significant discovery have been made in searching for pentaquark spectrum like $P_c$ or $P_{cs}$ structure.   Facing the present
status of the experiment in detecting the pentaquarks, more studies about pentaquarks, in theory, should be considered.
In fact, the hidden-charm pentaquarks $P_c$ and $P_{cs}$ were  predicted~\cite{Wu:2010vk,Lu:2016roh,Anisovich:2015zqa}, and suggested
to search for in different processes~\cite{Huang:2013mua,Feijoo:2015kts}.  We also note that the hidden-charm and hidden-bottom
pentaquark molecular states are studied by considering different details~\cite{Wang:2020bjt,Zhu:2020vto}.   In the current work,
we perform a study of possible pentaquark molecular states $\Lambda^0_b(5912)$ and $\Lambda^0_b(5920)$ productions in the $\gamma{}p\to{}\Lambda_b^{0(*)}B^+$ reactions.

In 2012, two narrow excited $\Lambda_b$ states, $\Lambda^0_b(5912)$ and $\Lambda^0_b(5920)$, were first observed by the the LHCb Collaboration
as a narrow peak in the $\Lambda_b^0\pi^{+}\pi^{-}$ invariant mass spectrum~\cite{Aaij:2012da}.  The latter state was confirmed by the CDF
Collaboration~\cite{Aaltonen:2013tta}. Following the discovery of the $\Lambda^0_b(5912)$ and $\Lambda^0_b(5920)$, many works treat them as
traditional bottom baryon~\cite{Wang:2017kfr,Gandhi:2020otj,Kawakami:2019hpp,Kawakami:2018olq}.  However, the pentaquark molecular states 
interpretation was also supported by Refs.~\cite{Romanets:2013cqa,Lu:2014ina,Liang:2014eba,GarciaRecio:2012db}.  Currently, 
High energy photon beams are available at Electron-Ion Collider in China (EicC)~\cite{Anderle:2021wcy}
or US(US-EIC)~\cite{Accardi:2012qut}, which provide another alternative to studying $\Lambda^0_b(5912)$ and $\Lambda^0_b(5920)$.   
Thus, it will be helpful to understand the nature of the $\Lambda^0_b(5912)$ and $\Lambda^0_b(5920)$ if we can observe them in
$\gamma{}p\to{}\Lambda_b^{0(*)}B^+$ production processes.

This paper is organized as follows. In Sec.~\ref{Sec: formulism}, we will
present the theoretical formalism. In Sec.~\ref{Sec: results}, the numerical
result of the $\gamma{}p\to{}\Lambda_b^{0(*)}B^{+}$ will be given, followed
by discussions and conclusions in last section.

\section{THEORETICAL FORMALISM}\label{Sec: formulism}
\begin{figure}[h!]
\begin{center}
\includegraphics[bb=100 510 750 710, clip,scale=0.65]{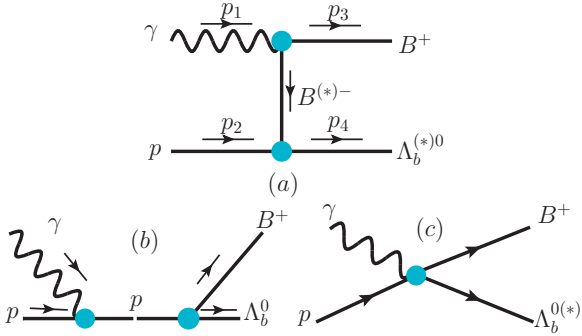}
\caption{Feynman diagrams for the $\gamma{}p\to{}\Lambda_b^{0(*)}B^{+}$ reaction. The contributions from the $t$-channel
$B^{(*)-}$ exchange(a),$s$-channel nucleon pole (b), and contact term (c) are considered. In
the first diagram, we also show the definition
of the kinematical $(p_1, p_2, p_3, p_4)$ that we use
in the present calculation.}
\label{cc}
\end{center}
\end{figure}

We study the $\gamma{}p\to{}\Lambda_b^{0(*)}B^{+}$ reactions within the effective Lagrangian approach, which has been widely
employed to investigate photoproduction processes.  The tree level Feynman diagrams for the $\gamma{}p\to{}\Lambda_b^{0(*)}B^{+}$
reactions are depicted in Fig.~\ref{cc}, where the contributions from the $t$-channel
$B^{-}$ and $B^{*-}$ exchange(a),$s$-channel nucleon pole (b), and contact term (c) are taken into account.

In order to computing the diagrams shown in Fig.~\ref{cc}, we need the
effective Lagrangian densities for the relevant interaction
vertices. As mentioned in the chiral unitary approach of Refs.~\cite{Romanets:2013cqa,Lu:2014ina,Liang:2014eba,GarciaRecio:2012db},
the $\Lambda^0_b(5912)$ and $\Lambda^0_b(5920)$ resonances are identified as $s$-wave
meson-baryon molecules that $\Lambda^0_b(5912)$ include big $N\bar{B}$ and $N\bar{B}^{*}$ components and $\Lambda^0_b(5920)$
possess big $N\bar{B}^{*}$ component.   For the $\Lambda^0_bN\bar{B}$, $\Lambda^0_bN\bar{B}^{*}$, and $\Lambda^{*0}_bN\bar{B}^{*}$
couplings, we can write down the Lagrangian densities as
\begin{align}
{\cal{L}}_{\Lambda^0_bN\bar{B}}&=g_{\Lambda^0_bN\bar{B}}\bar{\Lambda}_b^0N\bar{B},\\
{\cal{L}}_{\Lambda^0_bN\bar{B}^*}&=g_{\Lambda^0_bN\bar{B}^*}\bar{\Lambda}_b^0\gamma^{\mu}N\bar{B}^{*}_{\mu},\\
{\cal{L}}_{\Lambda^{0*}_bN\bar{B}^{*}}&=g_{\Lambda^{*0}_bN\bar{B}^{*}}\bar{\Lambda}_b^{*0\mu}N\bar{B}^{*}_{\mu},
\end{align}
The coupling constants in the above Lagrangians were determined in Ref.~\cite{GarciaRecio:2012db} in a
hadronic molecular picture with $g_{\Lambda^0_bN\bar{B}}=4.6$, $g_{\Lambda^0_bN\bar{B}^*}=3.0$, and $g_{\Lambda^{*0}_bN\bar{B}^{*}}=5.7$.

To compute the amplitudes of the diagrams shown in Fig.~\ref{cc},the effective Lagrangian densities related
to the photon fields are required, which are~\cite{Xiao:2017uve}
\begin{align}
{\cal{L}}_{\gamma{}pp}&=-e\bar{p}(\gamma^{\mu}-\frac{\kappa_p}{2m_p}\sigma^{\mu\nu}\partial_{\nu})A^{\mu}p,\\
{\cal{L}}_{\gamma{}B^*\bar{B}}&=\frac{g_{\gamma{}B^{*+}B^+}}{4}e\epsilon^{\mu\nu\alpha\beta}F_{\mu\nu}B^{*+}_{\alpha\beta}B^{-}\nonumber\\
                        &+\frac{g_{\gamma{}B^{*0}B^0}}{4}e\epsilon^{\mu\nu\alpha\beta}F_{\mu\nu}B^{*0}_{\alpha\beta}\bar{B}^{0}+H.c.,\label{eq5}\\
{\cal{L}}_{\gamma{}B^{-}B^{+}}&=ieA_{\mu}(B^{-}\partial^{\mu}B^{+}-B^{+}\partial^{\mu}B^{-}),
\end{align}
where the strength tensor is defined as $\sigma^{\mu\nu}=\frac{i}{2}(\gamma^{\mu}\gamma^{\nu}-\gamma^{\nu}\gamma^{\mu})$, $F_{\mu\nu}=\partial_{\mu}A_{\nu}-\partial_{\nu}A_{\mu}$ and $B^{*}_{\alpha\beta}=\partial_{\alpha}B^{*}_{\beta}-\partial_{\beta}B^{*}_{\alpha}$.
The anomalous magnetic momentum $\kappa_p=1.79$ and the $\alpha=e^2/4\pi=1/137$ is the electromagnetic fine structure constant.
The coupling constant $g_{\gamma{}B^{*+}B^+}$ and $g_{\gamma{}B^{*0}B^0}$ are determined from the partial decay
width of $B^{*+}\to{}B^{+}\gamma$ and $B^{*0}\to{}B^{0}\gamma$, which is obtained from Eq.~(\ref{eq5}),
\begin{align}
\Gamma(B^{*+}\to{}B^{+}\gamma)&=\frac{\alpha{}g^2_{\gamma{}B^{*+}B^+}}{24}m_{B^{*+}}(m^2_{B^{*+}}-m^2_{B^{+}}),\\
\Gamma(B^{*0}\to{}B^{0}\gamma)&=\frac{\alpha{}g^2_{\gamma{}B^{*0}B^0}}{24}m_{B^{*0}}(m^2_{B^{*0}}-m^2_{B^{0}}),
\end{align}
where the $m_{B^{*}}$ and $m_{B}$ are the mass of $B^{*}$ and $B$,respectively.  However, the width of the $B^{*}$
meson is not well determined experimentally. In the present work, we use the theoretical
predicted partial widths in Ref.~\cite{Choi:2007se} and the coupling
constants are determined to be $g_{\gamma{}B^{*+}B^+}=1.308$ GeV$^{-1}$ and $g_{\gamma{}B^{*0}B^0}=-0.745$ GeV$^{-1}$.

In evaluating the scattering amplitudes of the $\gamma{}p\to{}\Lambda_b^{0(*)}B^{+}$ reactions, we need to include the
form factors because hadrons are not pointlike particles.  For the $t$-channel $B$ and $B^{*}$ mesons exchange, we
would like to apply a widely used pole form factor, which is
\begin{align}
{\cal{F}}_{i}=\frac{\Lambda_i^2-m_i^2}{\Lambda_i^2-q_i^2}~~~~~~~ i=B,B^{*},
\end{align}
where $\Lambda_i=m_i+\alpha{}\Lambda_{QCD}$ and the QCD energy scale $\Lambda_{QCD}=220$ MeV.
The parameter $\alpha$ reflects the nonperturbative property of QCD at the low-energy scale,
which will be taken as a parameter and discussed later.  For the $s$-channel nucleon pole process,
we adopt a form factor
\begin{align}
{\cal{F}}_{N}(q^2,m^2_N)=\frac{\Lambda_N^4}{\Lambda_N^4+(q^2-m^2_N)^2}
\end{align}
with $\Lambda_N=0.9$ GeV, which can be well used to reproduce experimental
data of some reactions~\cite{Kim:2011rm,Oh:2004wp}.

With the above effective Lagrangian densities, the scattering amplitudes for the
$\gamma{}p\to{}\Lambda_b^{0}(5912)B^{+}$ and $\gamma{}p\to{}\Lambda_b^{0}(5920)B^{+}$
reactions can be obtained straightforwardly.  First, we write the scattering
amplitudes for the $\gamma{}p\to{}\Lambda_b^{0}(5912)B^{+}$ reaction
\begin{align}
{\cal{M}}^{B}_{t}&=-ieg_{\Lambda_b^0N\bar{B}}\bar{u}_{\Lambda_b^0}(p_4,s_4)u_p(p_2,s_2)\frac{1}{q_t^2-m^2_{\bar{B}}}(p_3^{\mu}-q_t^{\mu})\nonumber\\
                 &\times\epsilon_{\mu}(p_1,s_1){\cal{F}}_{B},\\
{\cal{M}}^{B^{*}}_{t}&=i\frac{eg_{\Lambda_b^0N\bar{B}^*}g_{\gamma{}B^{*+}B^+}}{4}\epsilon^{\rho\eta\alpha\beta}(p_{1\rho}g_{\eta\lambda}-p_{1\eta}g_{\rho\lambda})\nonumber\\
                     &\times(q_{t\alpha}g_{\beta\sigma}-q_{t\beta}g_{\alpha\sigma})\epsilon^{\lambda}(p_1,s_1)\frac{-g^{\mu\sigma}+q_t^{\mu}q_t^{\sigma}/m^2_{\bar{B}^{*}}}{q_t^2-m^2_{\bar{B}^*}}\nonumber\\
                     &\times{}\bar{u}_{\Lambda_b^0}(p_4,s_4)\gamma_{\mu}u_p(p_2,s_2){\cal{F}}_{B^{*}},\\
{\cal{M}}^{N}_{s}&=-ieg_{\Lambda_b^0N\bar{B}}\bar{u}_{\Lambda_b^0}(p_4,s_4)\frac{q\!\!\!/_s+m_p}{q^2_s-m_p^2}[\gamma^{\mu}+\frac{\kappa_p}{4m_p}(\gamma^{\mu}p\!\!\!/_1-p\!\!\!/_1\gamma^{\mu})]\nonumber\\
                 &\times{}u_p(p_2,s_2)\epsilon_{\mu}(p_1,s_1){\cal{F}}_{N}.
\end{align}
Then the amplitudes for the $\gamma{}p\to{}\Lambda_b^{0}(5920)B^{+}$ reaction have the form
\begin{align}
{\cal{M}}^{B^{*}}_{t2}&=i\frac{eg_{\Lambda_b^{*0}N\bar{B}^*}g_{\gamma{}B^{*+}B^+}}{4}\epsilon^{\rho\lambda\alpha\beta}(p_{1\rho}g_{\lambda\eta}-p_{1\lambda}g_{\rho\eta})\nonumber\\
                     &\times(q_{t\alpha}g_{\beta\nu}-q_{t\beta}g_{\alpha\nu})\epsilon^{\eta}(p_1,s_1)\frac{-g^{\mu\nu}+q_t^{\mu}q_t^{\nu}/m^2_{\bar{B}^{*}}}{q_t^2-m^2_{\bar{B}^*}}\nonumber\\
                     &\times{}\bar{u}^{\mu}_{\Lambda_b^{*0}}(p_4,s_4)u_p(p_2,s_2){\cal{F}}_{B^{*}}.
\end{align}
In the above equations, the $q_s=p_1+p_2=p_3+p_4$ and $q_t=p_1-p_3=p_4-p_2$.

The contact term illustrated in Fig.~\ref{cc}(c) serves to
keep the full amplitude gauge invariant. For the present
calculation, we adopt the following form
\begin{align}
{\cal{M}}^{c}_{\gamma{}p\to{}\Lambda_b^{0}(5912)B^{+}}&=ieg_{\Lambda_b^0N\bar{B}}\bar{u}_{\Lambda_b^0}(p_4,s_4)[{\cal{A}}(\gamma^{\mu}-1)+{\cal{B}}]\nonumber\\
                                                      &\times{}u_p(p_2,s_2)\epsilon^{\mu}(p_1,s_1),\\
{\cal{M}}^{c}_{\gamma{}p\to{}\Lambda_b^{0}(5920)B^{+}}&=0,
\end{align}
with
\begin{align}
{\cal{A}}=\frac{-m_p{\cal{F}}_{N}}{(q_s^2-m_p^2)}; {\cal{B}}=\frac{{\cal{F}}_{B}}{(q_t^2-m^2_{\bar{B}})}2p_3^{\mu}+\frac{{\cal{F}}_{N}}{q_s^2-m_p^2}2p_2^{\mu}.
\end{align}

The differential cross section in the center of mass
(c.m.) frame for the $\gamma{}p\to{}\Lambda_b^{0}(5912)B^{+}$ and $\gamma{}p\to{}\Lambda_b^{0}(5920)B^{+}$
reactions are calculated using the following equation:
\begin{align}
\frac{d\sigma}{d\cos\theta}=\frac{m_Nm_{\Lambda_b^{0(*)}}}{32\pi{}q^2_s}\frac{|\vec{p}^{c.m}_3|}{|\vec{p}^{c.m}_1|}\sum_{s_1,s_2,s_3,s_4}|{\cal{M}}_{1,2}|^2
\end{align}
where ${\cal{M}}_{1}={\cal{M}}^{B}_{t}+{\cal{M}}^{B^{*}}_{t}+{\cal{M}}^{N}_{s}+{\cal{M}}^{c}_{\gamma{}p\to{}\Lambda_b^{0}(5912)B^{+}}$
and ${\cal{M}}_{2}={\cal{M}}^{B^{*}}_{t2}$ are total scattering amplitude of the $\gamma{}p\to{}\Lambda_b^{0}(5912)B^{+}$ and $\gamma{}p\to{}\Lambda_b^{0}(5920)B^{+}$ reactions, respectively.
The $\theta$ is the scattering angle of the outgoing $B^{+}$ meson relative to the beam direction, while $\vec{p}^{c.m}_1$
and $\vec{p}^{c.m}_3$ are the photon and $B^{+}$ meson three momenta in the C.M. frame,respectively, which are
\begin{align}
|\vec{p}^{c.m}_1|=\frac{\lambda^{1/2}(q_s^2,0,m_N^2)}{2\sqrt{q^2_s}};~~~|\vec{p}^{c.m}_3|=\frac{\lambda^{1/2}(q_s^2,m^2_{B^{+}},m_{\Lambda_B^{0(*)}}^2)}{2\sqrt{q^2_s}}
\end{align}
where the $\lambda$ is the K\"{a}llen function with $\lambda(x,y,z)=(x-y-z)^2-4yz$.

\section{RESULTS}\label{Sec: results}
Considering the $\Lambda^0_b(5912)$ and $\Lambda^0_b(5920)$ as  pentaquark molecule,
their productions in the $\gamma{}p\to{}\Lambda_b^{0}(5912)B^{+}$
and $\gamma{}p\to{}\Lambda_b^{0}(5920)B^{+}$ reactions are evaluated.  The mechanism including
the $t-$ channel mediated by the exchange of $\bar{B}^{{*}}$ mesons, the contact term, and the $s-$ channel where
nucleon is considered an intermediate state.  To make a reliable prediction for the cross section of the
$\gamma{}p\to{}\Lambda_b^{0}(5912)B^{+}$ and $\gamma{}p\to{}\Lambda_b^{0}(5920)B^{+}$
reactions, the only issue we need to clarify is the explicit form of the parameter $\alpha$ relation to the form factors.
The parameter $\alpha$ reflects the nonperturbative property of QCD at the low-energy scale and could not be determined
by the first principles.  It is usually determined from the experimental branching ratios.  Next, how to
compute the value of $\alpha$ is shown in detail.

\begin{figure}[htbp]
\begin{center}
\includegraphics[bb=00 40 600 530, clip,scale=0.33]{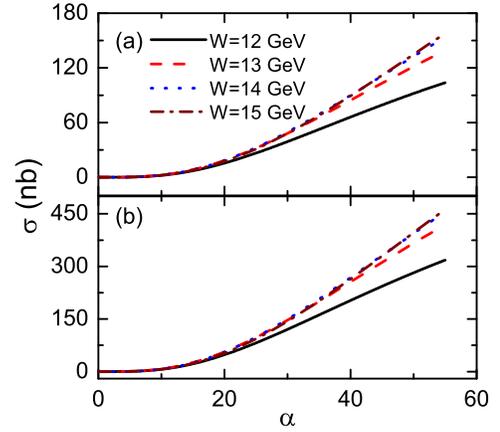}
\caption{(Color online)  Cross section of the $\gamma{}p\to{}\Lambda_b^{0}(5912)B^{+}$ (a) and the $\gamma{}p\to{}\Lambda_b^{0}(5920)B^{+}$ (b) with different energy
in the C.M. frame depending on the parameter $\alpha$.}
\label{decaywidth}
\end{center}
\end{figure}
With the formalism and ingredients give in Sec.~\ref{Sec: formulism}, the total cross section of the $\gamma{}p\to{}\Lambda_b^{0}(5912)B^{+}$
and $\gamma{}p\to{}\Lambda_b^{0}(5920)B^{+}$ reactions versus the model parameter $\alpha$ are computed.
The results obtained with several C.M energy $W$ are shown in Fig.~\ref{decaywidth}.  It finds that the value of the cross section
increases continuously but relatively slowly with the increasing
of $\alpha$ and in particular,  taking $W=12$ GeV as an example, a varying cutoff parameter from 0.0 to 5.0,
the value of the cross section is got from nearly $0.0$ to 0.2 nb and is not very sensitive to the model parameter $\alpha$ compared with
that cross section with lager $\alpha$.   This result is consistent with the findings
in Refs.~\cite{Xiao:2017uve,Dong:2017rmg,Chen:2012nva,Li:2012as,Colangelo:2002mj} that the $\alpha$ is restricted within a reasonable
range from $0.0$ to 5.0 by the experimental data.

\begin{figure}[htbp]
\begin{center}
\includegraphics[bb=00 00 600 410, clip,scale=0.35]{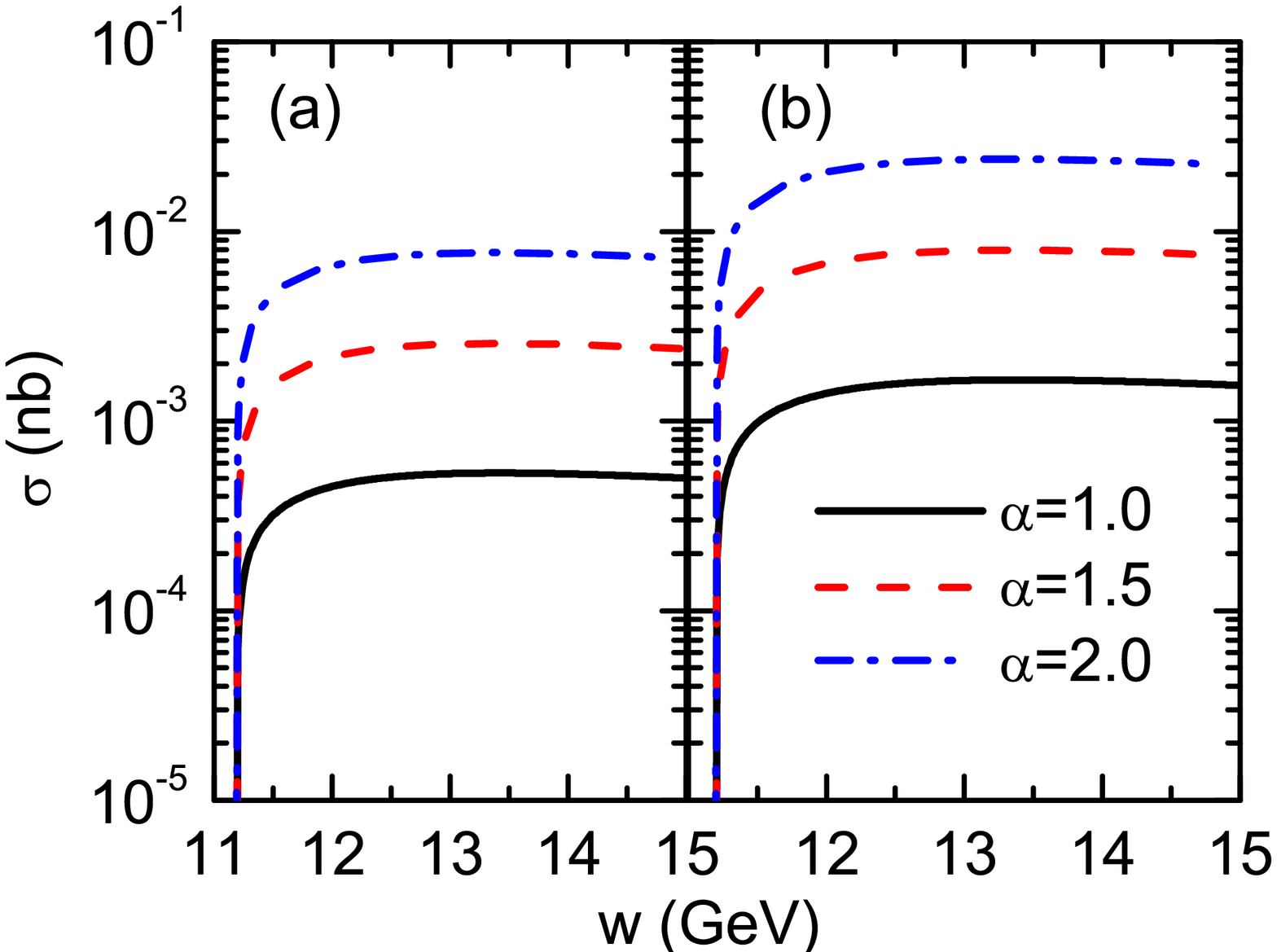}
\caption{(Color online) The total cross section for the processes (a) $\gamma{}p\to{}\Lambda_b^{0}(5912)B^{+}$ and (b)$\gamma{}p\to{}\Lambda_b^{0}(5920)B^{+}$
with different $\alpha$.}
\label{cross-1}
\end{center}
\end{figure}
Once the model parameter $\alpha$ are determined,  the total cross section versus the C.M energy W of the $\gamma{}p$ system
for the $\gamma{}p\to{}\Lambda_b^{0}(5912)B^{+}$ and $\gamma{}p\to{}\Lambda_b^{0}(5920)B^{+}$ reactions can be evaluated.
In Fig.~\ref{cross-1}, the total cross section of the $\gamma{}p\to{}\Lambda_b^{0}(5912)B^{+}$
and $\gamma{}p\to{}\Lambda_b^{0}(5920)B^{+}$ reactions with different $\alpha$ is presented , where we restrict
the $\alpha$ value within a reasonable range from 1.0 to 2.0.   It is worth mentioning that the value of the cross section
is very sensitive to the model parameter $\alpha$.
To see how much it depends on the cutoff parameter $\alpha$, we take the cross section at a energy about W=12 GeV with the
range of $\alpha=1.0-2.0$ as example.  The so-obtained the cross section
is got from $4.5\times{}10^{-4}$ nb to $6.6\times{}10^{-3}$ nb for $\gamma{}p\to{}\Lambda_b^{0}(5912)B^{+}$ reaction and
$1.4\times{}10^{-3}$ nb to $2.1\times{}10^{-2}$ nb for the $\gamma{}p\to{}\Lambda_b^{0}(5920)B^{+}$ reaction.
Such big change can not be suggested to search for $\Lambda_b^{0}(5912)$ and $\Lambda_b^{0}(5920)$
production in $\gamma{}p\to{}\Lambda_b^{0}(5912)B^{+}$ and $\gamma{}p\to{}\Lambda_b^{0}(5920)B^{+}$ reactions.
More accurate experimental data should be needed to constrain the value of the cutoff parameter $\alpha$.

\begin{figure}[htbp]
\begin{center}
\includegraphics[bb=50 00 700 410, clip,scale=0.33]{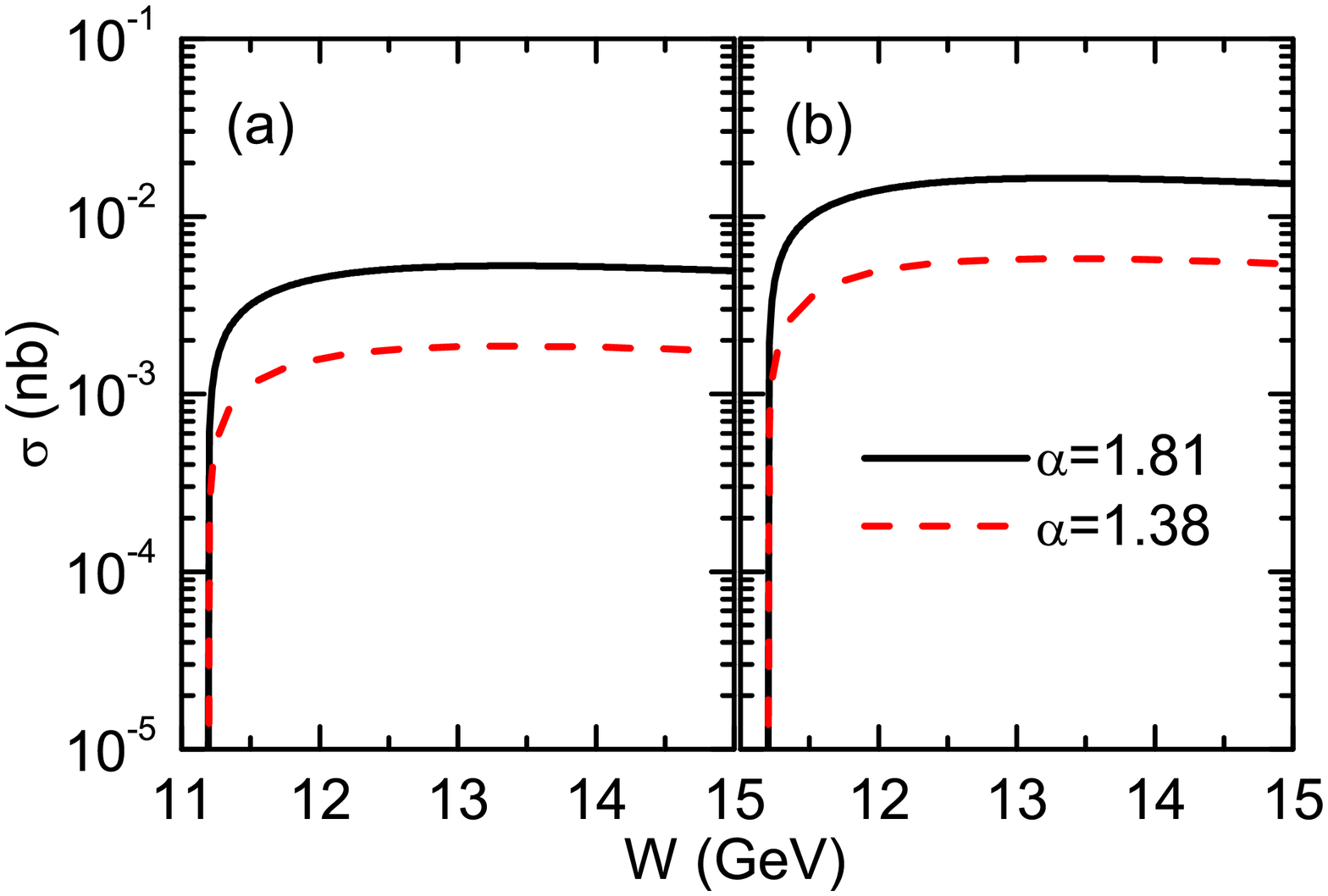}
\caption{(Color online) The total cross section for the processes (a) $\gamma{}p\to{}\Lambda_b^{0}(5912)B^{+}$ and (b)$\gamma{}p\to{}\Lambda_b^{0}(5920)B^{+}$
with different $\alpha$.}
\label{real-cross}
\end{center}
\end{figure}
Fortunately, more stringent constraints for the $\alpha$ value have been made by comparing with the experimental
data~\cite{Belle:2012koo,Colangelo:2002mj}.   As the free parameter in our calculation, $\alpha=1.38$ or $1.81$  is
fixed by fitting the experimental data of Ref.~\cite{Belle:2012koo},  whose procedures are just illustrated in
Ref.~\cite{Colangelo:2002mj}.  In this work, we adopt parameter $\alpha=1.38$ or $1.81$ because this value is determined
from the experimental data of Ref.~\cite{Belle:2012koo} within the same $B$ and $B^{*}$ form factors adopted in the
current work of Ref.~\cite{Colangelo:2002mj}.  The results for the C.M energy W from the reaction threshold to 15.0 GeV
are shown in Fig.~\ref{real-cross}.

From the Fig.~\ref{real-cross}, one can see that the total cross section increases sharply near the threshold.  At higher energies,
the cross section increases continuously but relatively slowly compared with that near threshold. However, the total cross section
decreases, but very slowly, when we change the C.M energy W form 13.6 GeV to 15.0 GeV.  The results also show that the total cross
section with $\alpha=1.81$ is bigger than that of $\alpha=1.38$. The difference between the cross section predicted
with  $\alpha=1.38$  and that predicted with $\alpha=1.83$ becomes larger with the energy increasing.

The results show that the total cross section for $\Lambda_b^{0}(5920)$ production is bigger than that for $\Lambda_b^{0}(5912)$ production.
At a C.M energy of about 13.6 GeV and a parameter $\alpha=1.81$($\alpha=1.38$), the cross section is of the order of 0.0164 (0.0057) nb for
$\Lambda_b^{0}(5920)$ production and 0.00527(0.00186) nb for $\Lambda_b^{0}(5912)$ production,  which is very challenging
to search for them at EicC~\cite{Anderle:2021wcy} but possible at US-EIC~\cite{Accardi:2012qut} due to a higher luminosity.
If the luminosity of EicC increases at least one order of magnitude, these states would be hopefully detected in photoproduction reaction.

\begin{figure}[htbp]
\begin{center}
\includegraphics[bb=50 20 700 520, clip,scale=0.28]{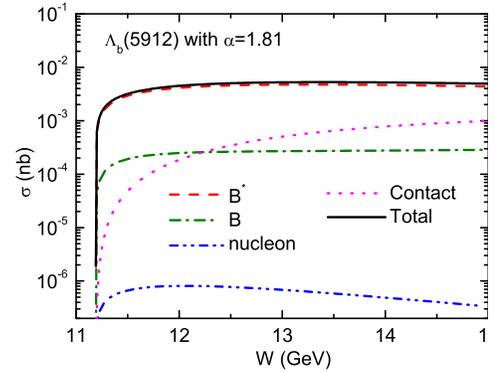}
\caption{(Color online)  Individual contributions of the the $t$-channel
$B^{}$(dash dot line) and $B^{*}$ (red dash line) exchange,$s$-channel nucleon pole (dash dot dot line), and contact term (dot line)
for the processes $\gamma{}p\to{}\Lambda_b^{0}(5912)B^{+}$ as a function of the energy.}
\label{contr-indi}
\end{center}
\end{figure}

We also find that the line shapes of the cross-section between the $\gamma{}p\to{}\Lambda_b^{0}(5912)B^{+}$ reaction and
the $\gamma{}p\to{}\Lambda_b^{0}(5920)B^{+}$ reaction are the same.  A possible explanation for this may be that the $t$-channel
$B^{*}$ meson exchange plays a predominant role in these two processes.   And this is guessed from the $\Lambda_b^0(5920)$
production mechanism that including only the $t-$ channel $\bar{B}^{*}$ meson exchange.  It indeed finds from Fig.~\ref{contr-indi} that
the individual contributions of the $t$-channel $\bar{B}^{*}$ meson exchange play a dominant role in the
$\gamma{}p\to{}\Lambda_b^{0}(5912)B^{+}$ reaction, while the contributions from the  $t$-channel $\bar{B}$ meson exchange, $s$-channel
nucleon pole and contact term are small.  Moreover, the interferences among them are quite small, which makes the $t$-channel
$\bar{B}^{*}$ meson exchange contribution almost saturates the total cross section.  The dominant $\bar{B}^{*}$ meson exchange contribution
can be easily understood since the $\Lambda_b^{0}(5912)$ and $\Lambda_b^{0}(5920)$ resonances are assumed as the molecular state
with a big $\bar{B}^{*}N$ component~\cite{Romanets:2013cqa,Lu:2014ina,Liang:2014eba,GarciaRecio:2012db}.

\begin{figure}[htbp]
\begin{center}
\includegraphics[bb=80 10 900 480, clip,scale=0.28]{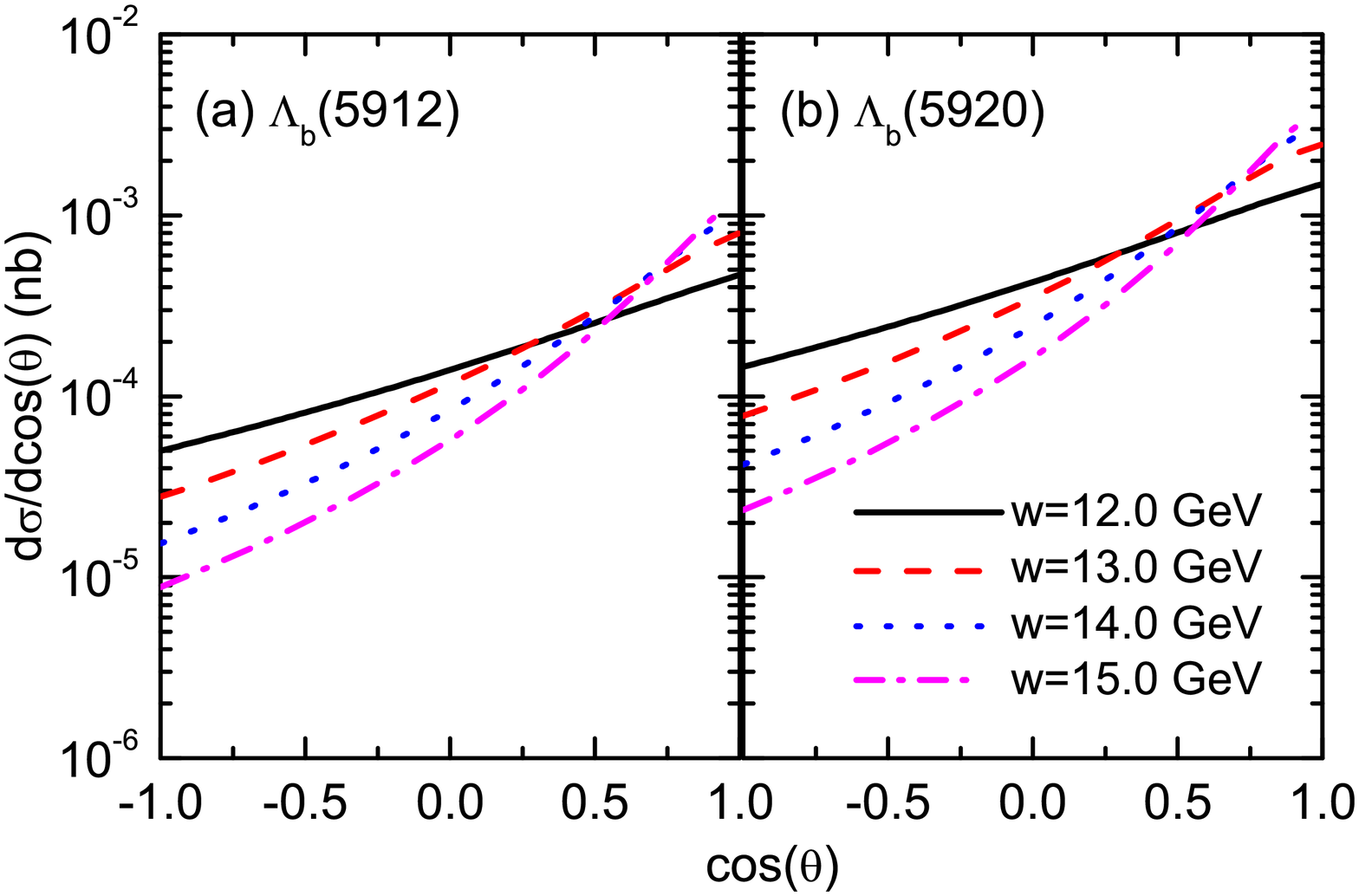}
\caption{(Color online)   (a)The $\gamma{}p\to{}\Lambda_b^{0}(5912)B^{+}$ and (b)$\gamma{}p\to{}\Lambda_b^{0}(5920)B^{+}$ differential
cross sections at different energies with $\alpha=1.81$. The black solid lines, red dashed lines, blue dotted lines, and
straight dashed-dotted lines are obtained at C.M energies W=12.0, 13.0, 14.0, and 15.0 GeV, respectively.}
\label{contr-dfff}
\end{center}
\end{figure}
In addition to the total cross section, we also compute the differential cross section for the $\gamma{}p\to{}\Lambda_b^{0}(5912)B^{+}$
and $\gamma{}p\to{}\Lambda_b^{0}(5920)B^{+}$ reactions as a function of the scattering angle of the outgoing meson relative to the beam direction at
different C.M energies, i.e., W=12.0, 13.0, 14.0, and 15.0 GeV.  The theoretical results are shown in Fig.~\ref{contr-dfff}. We note
that the differential cross section is the largest at the extreme forward angle and decreases with the increase of the scattering
angle.  This is because the the $t$-channel $\bar{B}^{*}$ meson exchange plays a predominant role in the $\gamma{}p\to{}\Lambda_b^{0}(5912)B^{+}$
and $\gamma{}p\to{}\Lambda_b^{0}(5920)B^{+}$ reactions.

\section{SUMMARY}
Stimulated by the newly observed pentaquark spectrum like $P_c$ or $P_{cs}$ structure,  pentaquark spectroscopy contain 
bottom quark is believed to be emerging.
In 2012, the $\Lambda^0_b(5912)$ and $\Lambda^0_b(5920)$ were first observed by the the LHCb Collaboration
as a narrow peak in the $\Lambda_b^0\pi^{+}\pi^{-}$ invariant mass spectrum~\cite{Aaij:2012da}. Many works treat them as
pentaquark molecular states~\cite{Romanets:2013cqa,Lu:2014ina,Liang:2014eba,GarciaRecio:2012db} with a $\bar{B}^{*}N$
component for $\Lambda^0_b(5920)$ and $\bar{B}^{(*)}N$ components for $\Lambda^0_b(5912)$. And they can not
be detected like $P_c$ or $P_{cs}$ through a heavy hadron decay to $\bar{B}^{(*)}N$ plus a light meson, since their masses
are just below the $\bar{B}^{(*)}N$ threshold.

In this paper, we made a detailed exploration of the nonresonant contribution to the $\gamma{}p\to{}\Lambda_b^{0}(5912)B^{+}$
and $\gamma{}p\to{}\Lambda_b^{0}(5920)B^{+}$, with the aim to find a reasonable estimation of the $\Lambda_b^{0}(5912)$ and
$\Lambda_b^{0}(5920)$ production rates at relatively high energies, where no data are available up to now.  The production
process is described by the $t$-channel $\bar{B}^{(*)-}$ exchange, $s$-channel nucleon pole, and contact term.  The coupling constants
of the $\Lambda^0_b(5912)$ to $\bar{B}^{(*)}N$ and $\Lambda^0_b(5920)$ to $\bar{B}^{*}N$ are obtained from chiral unitary
theory~\cite{Romanets:2013cqa,Lu:2014ina,Liang:2014eba,GarciaRecio:2012db}, where $\Lambda^0_b(5912)$ and $\Lambda^0_b(5920)$
are dynamically generated.

Our calculation indicates that the cross section for $\gamma{}p\to{}\Lambda_b^{0}(5912)B^{+}$ and $\gamma{}p\to{}\Lambda_b^{0}(5920)B^{+}$
reactions can reach 0.0164 nb and 0.00527 nb, respectively, are too small to be observed due to out of reach of the current luminosity design
$(2-4)\times{}10^{33}cm^{-2}s^{-1}$ of EicC~\cite{Anderle:2021wcy}.  If the luminosity of EicC increases at least one order of magnitude, 
these states would be hopefully detected in photoproduction reaction.  For the proposed Electron-Ion Collider in US (US-EIC) with the
luminosity of $10^{34}cm^{-2}s^{-1}$ or higher, it would be possible to observe these states~\cite{Accardi:2012qut}.  Moreover, the
differential cross sections computed also can be used to test the molecular picture of the $\Lambda^0_b(5912)$ and $\Lambda^0_b(5920)$.

\section*{Acknowledgments}
This work was supported by the Science and Technology
Research Program of Chongqing Municipal Education Commission (Grant No. KJQN201800510), the Opened Fund
of the State Key Laboratory on Integrated Optoelectronics
(GrantNo. IOSKL2017KF19). Yin Huang want to thanks
the support from the Development and Exchange Platform for
the Theoretic Physics of Southwest Jiaotong University under Grants No.11947404 and No.12047576,
the Fundamental Research Funds for the Central Universities(Grant No.
2682020CX70), and the National Natural Science Foundation
of China under Grant No.12005177.

\end{document}